\documentclass[12pt]{article}

\usepackage{amsthm,amsmath,amssymb,amsfonts} 

\usepackage[T1]{fontenc}
\usepackage[utf8]{inputenc}
\usepackage{authblk}

\def\s3{\sqrt{3}}
\def\s5{\sqrt{5}}

\newtheorem{Def}{Definition}
\newtheorem{The}{Theorem}

\newcommand{\bear}{\begin{array}}
\newcommand{\eear}{\end{array}}

\newfont{\tenbi}{cmbxti10}
\newcommand{\beq}{\begin{equation}}
\newcommand{\eeq}{\end{equation}}

\newcommand{\rd}{\mathrm{d}}

\newcommand\la{{\lambda}}

\newcommand\ka{{\kappa}}

\begin{document}

\title{Remarks on certain two-component systems with peakon solutions}
\author[1]{M. Hay}
\author[2]{Andrew N.W. Hone\footnote{On leave at School of Mathematics \& Statistics, 
UNSW, 
Australia.}}
\author[3]{Vladimir S. Novikov}
\author[2]{Jing Ping Wang}
\affil[1]{Dipartimento di Matematica e Fisica, Universit\`a di Roma Tre, Italy}
\affil[2]{SMSAS, 
University of Kent, UK}
\affil[3]{School of Mathematics, Loughborough University, UK}

\maketitle
\begin{abstract} 
We consider a Lax pair found by Xia, Qiao and Zhou for a family of   two-component   analogues of the Camassa-Holm equation, including an arbitrary function $H$, and show that this apparent freedom can be  removed via 
a combination of a reciprocal transformation and a gauge transformation, which reduces the system to triangular 
form. The resulting triangular system may or may not be integrable, depending on the choice of $H$. In addition, 
we apply the formal series  approach of Dubrovin and Zhang to show that scalar equations of Camassa-Holm type 
with homogeneous nonlinear terms of degree greater than three are not integrable.  
 \\ 
{\it This article is dedicated to Darryl Holm on his 70th birthday.} 
\end{abstract}
\section{Introduction} 

The  partial differential equation 
\beq\label{mch}
m_t+u m_x+2u_x m=0, \qquad m=u-u_{xx}
\eeq 
was derived  from asymptotic expansions in shallow water theory 
by Camassa and Holm \cite{CH}, who also obtained a 
bi-Hamiltonian structure and found remarkable weak solutions in the form 
of a superposition of peaks (peakons), 
\beq\label{peakons} 
u(x,t) = \sum_{j=1}^N p_j(t)\, e^{-|x-q_j(t)|}, 
\eeq 
where $(q_j,p_j)_{j=1,\ldots , N}$ form a set of canonical coordinates and momenta 
in a finite-dimensional Hamiltonian system with time $t$ that is completely integrable in the Liouville sense.  
Although it is a nonlocal partial differential equation, either in the form (\ref{mch}) in terms of $m$ with 
$u=(1-D_x^2)^{-1} \, m$, or rewritten as an evolution equation for $u$, i.e. $u_t = \ldots$ (cf. equation 
(\ref{ch}) below), the Camassa-Holm equation has an infinite hierarchy of commuting symmetries which 
are given by local evolution equations in $m$.

The 
integrability of the equation (\ref{mch}) itself was already included in earlier results 
of Fokas and Fuchssteiner on hereditary symmetries and recursion  operators \cite{FF}, but the work of 
Camassa and Holm led to  new analytical  and  geometrical insights: in addition   to the 
peakons given by (\ref{peakons}), and smooth solitons \cite{ach, matsuno, parker} that appear when linear 
dispersion is added to (\ref{mch}), other classes of initial data produce wave breaking \cite{mckean}; and 
the equation has a variational formulation as a geodesic flow on (an extension of) a diffeomorphism group 
\cite{mis}. The geometrical interpretation of (\ref{mch}) 
as  an Euler-Poincar\'e equation naturally generalizes to diffeomorphisms in two or more dimensions, and 
the analogues of the weak solutions (\ref{peakons}) can be applied to the problem of template matching in 
computational anatomy \cite{hmy}; but in general such higher-dimensional extensions do not  
to preserve integrability. Further research on the one-dimensional case has been concerned with the 
derivation \cite{fokas, or} and classification \cite{Novikov} of integrable scalar equations analogous to (\ref{mch}), 
as well as the search for suitable two-component or multi-component analogues 
\cite{CLZ, Falq, HI, hnw, strachan,  song, xiaq2, xiaq}. From the analytical point of view, 
there is also considerable interest in finding dispersive equations with higher order nonlinearity, 
which (despite not being integrable) display similar features in the form of peakons, wave breaking and one or more 
higher conservation laws \cite{anco, grayshan}.

Recently Xia, Qiao and Zhou introduced the following two-component system 
of partial differential equations:  
\beq \begin{array}{rcl} 
m_t& = & (mH)_x+mH-\frac{1}{2}m(u-u_x)(v+v_x), \\ 
n_t& = &\, (nH)_x\,-nH\,+\frac{1}{2}n(u-u_x)(v+v_x), \end{array} 
\label{syn} 
\eeq 
with 
\beq\label{mn} 
m=u-u_{xx}, \qquad n=v-v_{xx}. 
\eeq
In the above, $H$ is an arbitrary function of $x$ and $t$, which (in particular)  
can be fixed by choosing it to be a specific function 
of the fields $u,v$ and their derivatives. 
The authors of  \cite{xiaq} refer to this system in the title of their paper as ``synthetical'' 
because it provides a synthesis of several different 
systems admitting peakon solutions, by choosing $H$ to have a specific dependence on $u,v$; reductions to integrable 
scalar partial differential equations can be achieved by imposing further conditions on $u$ and $v$. For instance, 
setting $H=u$ and $v=2$ reduces (\ref{syn}) to the  Camassa--Holm equation \cite{CH}, which can be 
rewritten as 
\beq\label{ch}
(1-D_x^2)u_t=3uu_x-2u_xu_{xx}-uu_{xxx}; 
\eeq
while  setting $H=u^2-u_x^2$ and $v=2u$
produces the equation 
\beq\label{forq}
(1-D_x^2)u_t=D_x\left(u_x^2u_{xx}-u^2u_{xx}-uu_x^2+u^3\right),
\eeq 
which was first derived by Fokas \cite{fokas}, then by Olver and Rosenau \cite{or}, and has been studied more recently by Qiao \cite{Qiao}.  
Other examples include the two-component system 
\beq \begin{array}{rcl} 
m_t& = & \frac{1}{2}D_x\Big(m(u-u_x)(v+v_x)\Big), \\ 
n_t& = &\, \frac{1}{2}D_x\Big(n(u-u_x)(v+v_x)\Big)\end{array} 
\label{song} 
\eeq 
introduced in \cite{song}, whose multipeakon solutions were recently analyzed in \cite{chang}, which 
arises  from (\ref{syn}) by taking 
\beq\label{songh} 
H=\frac{1}{2}(u-u_x)(v+v_x); 
\eeq 
or the system obtained from the choice 
\beq\label{xiah} 
H=\frac{1}{2}(uv-u_xv_x),
\eeq 
 that is 
\beq \begin{array}{rcl} 
m_t& = & \frac{1}{2}D_x\Big(m(uv-u_xv_x)\Big)-\frac{1}{2}\Big(m(uv_x-u_xv)\Big), \\ 
n_t& = &\, \frac{1}{2}D_x\Big(n(uv-u_xv_x)\Big)+\frac{1}{2}\Big(n(uv_x-u_xv)\Big),\end{array} 
\label{xiaq} 
\eeq 
which was studied in \cite{xiaq2}. 

However, the title of the paper \cite{xiaq} is somewhat unfortunate, since in English the word ``synthetic'' is more commonly used, and  it is synonymous with ``artificial'' or ``fake''  in  everyday  language. 
Although the system (\ref{syn}) arises as the 
compatibility condition of a linear system (Lax pair), which yields an infinite sequence of conservation laws, we will show 
that this is not sufficient for this two-component system (or all its reductions) to be integrable. In fact, the linear system obtained in   
\cite{xiaq} should be considered as an example of a fake Lax pair (see \cite{fakecn} or \cite{butlerhay,sako}): by a combination of   
a change of independent variables (reciprocal transformation) and a gauge 
transformation, the function $H$ can be removed, 
and  the system decouples into an integrable scalar equation together with an arbitrary evolution equation (which is  
generically non-integrable). As the consequence, the infinite sequence of conservation laws only depend on a single dependent variable (the variable $\vartheta$ below).
Thus it  turns out that it is  appropriate
to apply the word ``synthetic'' in this context. 

Our main result can be summarized as follows. 

\begin{The}\label{main1} Let (\ref{syn}) be specified as an autonomous system 
of partial differential equations for $u=u(x,t)$ and $v=v(x,t)$, by making  
a particular choice of function 
$H=h(u,v,u_x,v_x,\ldots)$ of $u,v$ and their $x$-derivatives. Then there is a reciprocal transformation 
to a triangular system for $\vartheta=\vartheta(X,T)$ and 
$\ka = \ka (X,T)$, given by  
\beq \begin{array}{rcl} 
\vartheta_T+\frac{1}{4}\left( \frac{(\vartheta_{XT}-2)^2}{4\vartheta^2}-\vartheta_T^2\right)_X & = &  0, \\ 
\ka_T +\ka \, {\cal F}[\vartheta , \ka ] & = & 0
,\end{array} 
\label{triang} 
\eeq 
where ${\cal F} [\vartheta , \ka ]$ denotes a (possibly nonlocal) function of $\vartheta$, $\ka$ and their $X$-derivatives. 
\end{The}

As we shall see, on its own the first equation for $\vartheta$ in the system (\ref{triang}) is integrable: it corresponds to a 
negative flow in the modified KdV hierarchy; but in general the second equation is not integrable, for an arbitrary choice of $\cal F$ (which corresponds to the arbitrariness of $H$). In the next section we will derive the above result, 
explaining how the dependence of the system (\ref{syn}) on $H$ can be removed from the Lax pair. 

Section 3 is 
concerned with a different question, namely the degree of nonlinearity that appears in integrable peakon equations. 
Using the approach of Dubrovin and Zhang, which is based on writing equations as series that are perturbations of the dispersionless 
limit \cite{dz1, dz2, dz3}, we present a theorem to the effect that there are no homogeneous scalar 
peakon equations with nonlinearity of degree greater than three. 
This result should be sufficient to infer that all integrable multi-component analogues of the 
Camassa-Holm can have only quadratic or cubic nonlinear terms, since such systems 
reduce to the scalar case by setting all but one of the fields to zero. The paper ends with a brief discussion of the results.

\section{Lax pair and reciprocal transformation} 

For what follows it will be convenient to rescale the dependent variables $u,v$ in (\ref{syn}) so that 
$$ 
u\rightarrow 2u, \qquad v\rightarrow -2v, 
$$ 
which implies that $m\to 2m$, $n\to-2n$, and introduce the quantities 
\beq\label{ABmiu} 
A=u-u_x, \qquad B=v+v_x, 
\eeq 
so that the system takes the form 
\beq \begin{array}{rcl} 
m_t& = & (mH)_x+mH+2mAB , \\ 
n_t& = &\, (nH)_x\,-nH\,-2nAB, \end{array} 
\label{ssyn} 
\eeq 
with 
\beq\label{AB} 
m=A+A_{x}, \qquad n=B-B_{x}. 
\eeq
With this choice of scaling, the Lax pair presented in (\ref{xiaq}) can be rewritten as 
\beq\label{laxp} 
\Psi_x={\bf U}\Psi, \qquad \Psi_t={\bf V}\Psi, 
\eeq 
where 
$$ 
{\bf U}=\left(\begin{array}{cc} -\frac{1}{2} & m\la \\ 
n\la & \frac{1}{2}
\end{array}\right), \qquad 
{\bf V}=\left(\begin{array}{cc}-\frac{1}{2}\la^{-2}+AB & A\la^{-1}+mH\la \\ 
  B\la^{-1}+nH\la & \frac{1}{2}\la^{-2}-AB 
\end{array}\right)
.
$$ 

By a standard method, transforming the $x$ part of (\ref{laxp}) into a Riccati equation and making an asymptotic expansion 
of the Riccati potential in powers of $\la$, it was shown in \cite{xiaq} that the system (\ref{syn}) has infinitely many conservation laws. 
With the choice of  scaling as in (\ref{ssyn}),  the first of 
these is 
\beq\label{cons} 
q_t=(qH)_x, \qquad q=\sqrt{mn}. 
\eeq 
Using the latter, we transform the independent variables via the reciprocal transformation 
\beq\label{rt} 
\rd X= q\,\rd x +qH\,\rd t, \qquad \rd T=\rd t, 
\eeq 
so that partial derivatives transform as 
$D_x =q\, D_X$, $D_t=D_T+qH\, D_X$. It is then helpful to replace $m,n$ throughout by $q,\ka$, where  
$$ 
\ka =\sqrt{\frac{n}{m}}, 
$$  
so that the system (\ref{ssyn}) becomes 
\beq\label{pk} 
\begin{array}{rcl}
(q^{-1})_T+H_X & = & 0, \\ 
(\log \ka)_T+H+2AB & = & 0, 
\end{array} 
\eeq 
and (\ref{AB}) produces 
\beq\label{ABX} 
A_X=-Aq^{-1}+\ka^{-1}, \qquad B_X=Bq^{-1}-\ka. 
\eeq 
The reciprocal transformation can also be applied to the Lax pair (\ref{laxp}), to yield 
\beq\label{laxprt} 
\Psi_X=\hat{\bf U}\Psi, \qquad \Psi_T=\hat{\bf V}\Psi, 
\eeq 
where 
$$ 
\hat{\bf U}=\left(\begin{array}{cc} -\frac{1}{2}q^{-1} & \ka^{-1}\la \\ 
\ka\la & \frac{1}{2}q^{-1}
\end{array}\right), \,
\hat{\bf V}=\left(\begin{array}{cc}-\frac{1}{2}\la^{-2}+AB+\frac{1}{2}H & A\la^{-1}\\ 
  B\la^{-1} & \frac{1}{2}\la^{-2}-AB -\frac{1}{2}H
\end{array}\right)
.
$$  

The compatibility conditions for (\ref{laxprt}), 
coming from the zero curvature equation 
$
{\bf U}_t -{\bf V}_x + [{\bf U}, {\bf V}]=0$,
 are precisely the equations (\ref{pk}) and (\ref{ABX}).
We  now explain how these equations  can be decoupled into a triangular system consisting of an integrable scalar equation  
together with an arbitrary evolution equation (which is generically non-integrable). To see this, we introduce the gauge 
transformation 
$$ 
\Psi = {\bf g} \Phi, \qquad {\bf g} =\left(\begin{array}{cc} \ka^{-1/2} & 0 \\ 0 & \ka^{1/2} \end{array}\right), 
$$ 
which transforms the Lax pair (\ref{laxprt}) to 
\beq\label{laxpint} 
\Phi_X={\cal U}\Phi, \qquad \Phi_T={\cal V}\Phi, 
\eeq 
where 
$$ 
{\cal U}=\left(\begin{array}{cc} \vartheta & \la \\ 
\la & -\vartheta
\end{array}\right), \quad 
{\cal V}=\left(\begin{array}{cc}-\frac{1}{2}\la^{-2}& F\la^{-1}\\ 
  G\la^{-1} & \frac{1}{2}\la^{-2} 
\end{array}\right)
,
$$  
with 
\beq\label{thdef} 
\vartheta=-\frac{1}{2q}+\frac{1}{2}(\log\ka)_X, \qquad F=\ka A, \qquad G=\ka^{-1}B.
\eeq 
The compatibility conditions for (\ref{laxpint}) are 
\beq\label{ints}
\begin{array}{rcl} 
\vartheta_T & = & F-G, \\ 
F_X & = & 2\vartheta F+1, \\ 
G_X & = & -2\vartheta G-1.   
\end{array} 
\eeq 
The latter system can be written as a single scalar  equation for $\vartheta$, by eliminating $F$ and $G$. This is best achieved by noting that the second and third equations in (\ref{ints}) imply 
$ 
(FG)_X=G-F
$, and so the first equation yields the conservation law 
\beq\label{consl} 
\vartheta_T +(FG)_X=0. 
\eeq 
The difference of the last two equations     in (\ref{ints}) also gives 
$
2\vartheta (F+G)=(F-G)_X-2=\vartheta_{XT}-2
$ 
(using the first equation once again), so that overall we have $F$ and $G$ given in terms of $\vartheta$ as 
\beq\label{FG} 
F=\frac{\vartheta_{XT}-2}{4\vartheta} +\frac{1}{2}\vartheta_T, \qquad 
G=\frac{\vartheta_{XT}-2}{4\vartheta} -\frac{1}{2}\vartheta_T. 
\eeq 
Finally, if we substitute these expressions into the conservation law (\ref{consl}), we obtain the 
equation 
\beq\label{negmkdv} 
\vartheta_T+\frac{1}{4}\left( \frac{(\vartheta_{XT}-2)^2}{4\vartheta^2}-\vartheta_T^2\right)_X=0, 
\eeq  
which is an integrable partial differential equation for $\vartheta(X,T)$. In fact, (\ref{negmkdv}) corresponds to the first negative flow in the modified KdV hierarchy, which is best seen by rewriting (\ref{laxpint}) as a scalar Lax pair for the 
first component $\phi=\phi_1$ of the vector $\Phi=(\phi_1,\phi_2)^T$, to find 
$$ 
\begin{array}{rcl} 
(D_X+\vartheta)(D_X-\vartheta)\phi & \equiv & (D_X^2+W)\phi=\la^2\phi, \\ 
\phi_T & =& \la^{-2}\Big(F\phi_X-\frac{1}{2}F_X\phi\Big), 
\end{array} 
$$  
with 
$$
W=-\vartheta_X-\vartheta^2; 
$$ 
the $X$ part is just the KdV spectral problem, and the potential $W$ in the Schr\"{o}dinger operator is 
given by the standard Miura expression in terms of $\vartheta$. 
However, observe that the time evolution 
of the field $\ka(X,T)$ is not determined by the Lax pair (\ref{laxpint}); we shall return to this shortly.  

Since (\ref{negmkdv}) takes the form of a conservation law, it is convenient to introduce a potential $f(X,T)$ such 
that $\vartheta=f_X$, and then the equation 
\beq\label{feq} 
f_T +\frac{(f_{XXT}-2)^2}{16f_X^2}-\frac{f_{XT}^2}{4}=0 
\eeq 
is obtained, by integrating and absorbing an arbitrary function of $T$ into $f$. Thus we can describe  solutions of the 
system (\ref{ssyn}) in the following way. 
\begin{The}\label{main} 
Let $f(X,T)$ be  a solution of (\ref{feq}), let $\ka(X,T)$ be an arbitrary function, and let $\vartheta(X,T)=f_X(X,T)$. 
 Then a solution $(A(x,t),B(x,t))$ of the system (\ref{ssyn}) with non-autonomous coefficient $H(x,t)$ is 
given in parametric form by setting 
$$ 
x=\log\ka (X,t)-2f(X,t) 
$$ 
and $T=t$ in the expressions 
$$ 
A=\ka^{-1}F, \qquad B=\ka G, \qquad H=-(\log\ka)_T-2FG, 
$$ 
where $F$ and $G$ are given in terms of $\vartheta$ by (\ref{FG}). 
\end{The}

In the above formulation, the function $\ka$ is arbitrary, and together with $\vartheta$ it completely determines the 
quantity  $H$, viewed as a non-autonomous coefficient appearing in the system  (\ref{ssyn}). However, in \cite{xiaq} the role of $H$ was envisaged somewhat differently: in that setting, one must consider the inverse problem of determining $\ka (X,T)$, when $H$ is some given function of the original fields and their derivatives that is specified a priori, i.e. $H=h(u,v,u_x,v_x,u_{xx},v_{xx},\ldots)$ in (\ref{syn}), or $H=\hat{h}(A,B,A_x,B_x,\ldots)$ in (\ref{ssyn}). With this alternative perspective, the original coupled system for $u,v$ (or $A,B$) with independent variables $x,t$ is equivalent under the reciprocal transformation 
(\ref{rt})  to a system consisting of the integrable equation (\ref{negmkdv}) together with the evolution equation 
\beq\label{kaevol} 
\ka_T = -\ka (H+2FG) 
\eeq
for $\ka$, with $F$ and $G$ given by (\ref{FG}). The latter system is triangular, since  (\ref{negmkdv}) is an 
autonomous equation for $\vartheta$ alone, while the terms on the right-hand side of (\ref{kaevol}) generally depend on 
both $\ka$, $\vartheta$ and their derivatives in a complicated way; for instance, given  $H=\hat{h}(A,B,A_x,B_x,\ldots)$ 
we should replace $A,B$ by $A=\ka^{-1}F$, $B=\ka G$ and use  (\ref{FG}), while $A_x$ should be replaced by 
$qA_X=q(\ka^{-1}F)_X$ where 
$$q=\Big((\log\ka)_X-2\vartheta\Big)^{-1}, 
$$ 
and so on; alternatively, given $H=h(u,v,u_x,v_x,\ldots)$ one must consider potentially nonlocal 
expressions, since  (\ref{ABmiu}) gives $u-u_x=A$, so $u-qu_X= \ka^{-1}F$, etc. Moreover, the equation 
(\ref{kaevol}) is completely independent of the Lax pair (\ref{laxpint}), so there is no reason for it to be 
integrable. 

Thus, by identifying ${\cal F}=H+2FG$,  we have arrived at Theorem \ref{main1}, and our main conclusion: in general, for a given choice of $H$, the Lax pair 
(\ref{laxp}) is insufficient to infer that the system (\ref{syn}) is integrable.  An integrable coupled system only arises 
for certain exceptional choices of $H$. 

One particular exception is the case corresponding to (\ref{songh}) above, namely 
$$ 
H=-2AB, 
$$ 
which causes the right-hand side of (\ref{kaevol}) to vanish, since $FG=AB$. In that case, in terms of $A,B$ 
with $m,n$ given by (\ref{AB}), the system 
(\ref{ssyn}) takes the form   
\beq\label{cub2}
m_t=-2(ABm)_x, \qquad n_t=-2(ABn)_x, 
\eeq
which is one of the coupled cubic integrable systems derived recently in \cite{hnw}.  After rescaling the dependent variables, this corresponds to the system (\ref{song}) obtained in \cite{song}, via the Miura map (\ref{ABmiu}); the equation 
(\ref{forq}) is a reduction of this system to a scalar equation. This is a situation where the equation (\ref{kaevol})  trivially 
decouples from (\ref{negmkdv}), with $\ka_T=0$ implying that $\ka$ is an arbitrary function of $X$.  For 
some exact solutions  of the system (\ref{cub2}), see \cite{hnw}. 

Another exceptional situation arises by taking a reduction to a scalar equation with 
$$ 
H=ku, \qquad v=\ell, \quad \text{for}\, \,k,\ell\, \,\text{constant}, 
$$ 
so that $B=n=\ell$. 
From the second equation in (\ref{ssyn}) it follows that 
$\ell=-k/2$ must hold, and the first equation becomes the Camassa-Holm equation (\ref{ch}), up to rescaling. To fix the choice of scale we set $k=-2$, $\ell=1$, to find  
$$ 
q=\sqrt{m}=\ka^{-1}, 
$$ 
while using $FG=AB$ gives 
$$ 
H+2FG=-2u+2(u-u_x)=-2qu_X=qH_X, 
$$ 
from which it follows that the two equations in (\ref{pk}) are equivalent to each other, and by 
(\ref{thdef}) we see that $\vartheta$ is given in terms of $q$ as 
$$ 
\vartheta=\frac{q_X-1}{2q}. 
$$ 
The field $q$ satisfies the equation 
$$ 
(q^{-1})_T + \Big(q(\log q )_{XT}-2q^2\Big)_X=0, 
$$ 
which can be rewritten in the form 
$$W_T=-2q_X, \qquad W=-\frac{q_{XX}}{2q}+\frac{q_X^2}{4q^2}-\frac{1}{4q^2}, $$ 
identifying it as the first negative flow in the KdV hierarchy (see \cite{Fuchss,WH} for more details). 

However, for other choices of $H$ we expect that an integrable system does not arise; in particular, it appears 
that  
the system (\ref{xiaq}) and other examples considered in \cite{xiaq} are not integrable.  

\section{Homogeneous Camassa-Holm type equations}

In this section we consider integrable Camassa-Holm type equations 
with  homogeneous nonlinear terms, of the form 
\begin{equation}
\label{chhom}
(1-D_x^2)u_t=\alpha u^k u_x+\beta u^k u_{xxx}+\gamma u^{k-1}u_xu_{xx}+\delta u^{k-2}u_x^3.
\end{equation}
Nonlinearities of this type have been considered in \cite{anco, grayshan}. 
Here $\alpha,\beta,\gamma,\delta, k$ are arbitrary complex constants, and we assume that
$\alpha \, k\neq 0$. 

Known integrable examples of Camassa-Holm type equations of the form (\ref{chhom}) correspond to $k=1$ and $k=2$. 
We show that under the above assumptions these are the only possible degrees of nonlinearity.

To prove this we adopt the Dubrovin-Zhang approach \cite{dz1, dz2, dz3}. Consider a formal series
\begin{equation}
\label{formeq}
\begin{array}{rcl}
u_t & = & \lambda(u)u_x+\epsilon\left(a_1(u)u_{xx}+a_2(u)u_x^2\right) 
\\ 
&&  +\, \epsilon^2\left(b_1(u)u_{xxx}+b_2(u)u_xu_{xx}+b_3(u)u_x^3\right)+\cdots , 
\end{array} 
\end{equation}
where $\epsilon$ is an arbitrary parameter, and we assume that $\lambda'(u)\ne 0$. Expressions at each 
power $\epsilon^n$ are homogeneous differential polynomials in $x$-derivatives of $u$ of weight $n+1$ if we adopt the convention that the weight of $u$ is $0$ and the weight of the $j$th derivative of $u$ is $j$, $j\ge 0$.  The series (\ref{formeq}) may or may not truncate. In the former case the expression (\ref{formeq}) is an evolutionary partial differential equation. Following Dubrovin-Zhang we adopt the following definition of {\it integrability} for the formal series (\ref{formeq}):

\begin{Def} The series (\ref{formeq}) is integrable if there exists another formal series 
\begin{equation}
\label{formsym}
\begin{array}{rcl}
u_{\tau} &= &\mu(u)u_x+\epsilon\left(A_1(u)u_{xx}+A_2(u)u_x^2\right)  \\ 
&& +\epsilon^2\left(B_1(u)u_{xxx}+B_2(u)u_xu_{xx}+B_3(u)u_x^3\right)+\cdots  
\end{array} 
\end{equation}
that commutes with (\ref{formeq}) for an {\it arbitrary} choice of the function $\mu(u)$.
\end{Def}
 
Taking the dispersionless limit $\epsilon\to 0$ in  (\ref{formeq}) yields 
an equation in the (dispersionless)  
Burgers hierarchy, which has  $u_{\tau}=\mu(u)u_x$ as a symmetry for any $\mu(u)$. 
The above definition suggests the following integrability test for (\ref{formeq}):
\begin{itemize}
\item[\bf{1.}] The commutator of (\ref{formeq}) and (\ref{formsym}), if nonzero, is a series in positive powers of $\epsilon$. At every power of epsilon we compute coefficients at every monomial $D_x^{n_1}(u)\cdots D_x^{n_k}(u)$, $n_1,n_2,\ldots >0$.
\item[\bf{2.}] We require that coefficients at {\it every} monomial $(\mu^{k_1}(u))(\mu'(u))^{k_2}(\mu''(u))^{k_3}\cdots$ vanish. These form a system of ordinary differential equations for the functions $\lambda(u),a_1(u),a_2(u),b_1(u),\ldots$, which are the integrability conditions for (\ref{formeq}).
\end{itemize}

The application of the above test to the Camassa-Holm type equations (\ref{chhom}) 
with  homogeneous nonlinearity is the following. First of all we rewrite (\ref{chhom}) as an evolutionary formal series by inverting the operator $1-D_x^2$:
\beq
\label{chhomser}
\begin{array}{rcl} 
u_t&=&(1-D_x^2)^{-1}(\alpha u^k u_x+\beta u^k u_{xxx}+\gamma u^{k-1}u_xu_{xx}+\delta u^{k-2}u_x^3)\\ 
&=& 
\alpha u^ku_x+(\alpha+\beta)u^ku_{xxx}+(3k\alpha+\gamma)u^{k-1}u_xu_{xx} \\
&&
+(\delta+k(k-1)\alpha)u_x^3+\cdots.
\end{array} 
\eeq
By rescaling $x$ and $t$ it is convenient to introduce the  parameter $\epsilon$, which counts the weight of every monomial:
\begin{equation}
\label{chhomser1}
u_t =  \alpha u^ku_x  
+\epsilon^2\, F_2[u] + 
\epsilon^4\, F_4[u]+\cdots, 
\end{equation}
where 
$$  F_2[u] = 
(\alpha+\beta)u^ku_{xxx}+(3k\alpha+\gamma)u^{k-1}u_xu_{xx}
+(\delta+k(k-1)\alpha)u_x^3 
$$ 
and 
the omitted terms are $O(\epsilon^6)$. 
(It is easy to see that all odd orders of $\epsilon$ 
are absent from the above expression.)

The application of the above test to the formal series (\ref{chhomser1}) leads to the following theorem:

\begin{The}
If equation (\ref{chhom}) with $\alpha\ne 0$ and $k\ne 0$ is integrable then $k=1$ or $k=2$.
\end{The}

The condition that $\alpha k\neq 0$ in the theorem guarantees the applicability of the test. The proof requires computations up to the order $\epsilon^{10}$. The resulting set of conditions is an algebraic system of equations for $\alpha,\beta,\gamma,\delta$ and $k$, which we do not present here due it  being too cumbersome. This system possesses nontrivial (nonzero) solutions only if $k=1$ or $k=2$.

\section{Discussion} 

Lax pairs and an infinite number of conservation laws are considered to be 
 hallmarks of integrability for systems of partial differential equations.
However,  a rigorous definition of these concepts is required, in particular 
for multi-component systems. Without it,
a Lax pair alone is not sufficient to infer integrability \cite{butlerhay, fakecn, sako}, and 
even an infinite number of conservation laws may not be enough.  No matter what 
choice of $H$ is made, the system (\ref{syn}) formally arises from a Lax pair, and this 
Lax pair yields an infinite number of conservation laws, but our calculations show 
that this Lax pair is ``fake'' in the sense that the dependence on $H$ can be removed, 
and the system can be reduced to an integrable scalar equation coupled with another 
equation which is not integrable in general. Nevertheless, there are certain specific choices of 
$H$ for which the second equation is either trivial or equivalent to a copy of the first equation, 
corresponding to the Camassa-Holm equation, or to the system (\ref{song}) found in \cite{song} (which includes 
(\ref{forq}) as  a scalar reduction).

There are  other systems  found in \cite{xiaq} for which compatible bi-Hamiltonian operators are 
presented, including the system (\ref{xiaq}) from \cite{xiaq2}. As bi-Hamiltonian 
structures are considered to be another hallmark of integrability, this would seem to contradict 
our claim that these other choices of $H$ should not give integrable systems. However, it 
appears that the pairs   of compatible Hamiltonian operators $J,K$ presented for these other cases 
in \cite{xiaq} do not give rise to an infinite hierarchy of {\it local} symmetries in terms of 
the fields $m,n$. Within the symmetry approach to integrable systems \cite{MSS, MN, review}, there is a 
requirement of infinitely many local symmetries, yet most of the recursion operators 
$JK^{-1}$ or $KJ^{-1}$ found in \cite{xiaq} produce only nonlocal equations, so there 
is no contradiction. 

Furthermore, we expect that in the peakon sector  it may not possible be obtain a consistent spectral 
problem from the Lax pair for (\ref{xiaq}) and the other systems presented in  \cite{xiaq}, 
apart from the exceptional system (\ref{song}), for which the spectral theory for the peakons 
was derived in \cite{chang}.


All of the known integrable scalar equations or coupled systems of Camassa-Holm type contain 
nonlinear terms of degree at most three. The result in the third section above shows that this condition 
on the degree is necessary for integrability. 

\noindent 
{\bf Acknowledgments:} 
MH and ANWH both benefited  from the hospitality provided by the Dipartimento di Matematica e Fisica 
di Roma Tre  in 2013, when we began discussions on fake Lax pairs. ANWH would like to thank the organizers 
of New Trends in Applied Geometric Mechanics at ICMAT (Madrid, Spain) in July 2017 for such a wonderful 
celebration  of Darryl Holm's 70th birthday. We are all grateful to Darryl for so much warm friendship and 
inspiration over the years.

\end{document}